\documentclass[prd, amsfonts, twocolumn, nofootinbib, showpacs]{revtex4-2}
\usepackage{graphicx, epsfig}
\usepackage{color}
\usepackage{amsmath}
\usepackage{amssymb}
\usepackage{physics}
\newcommand{\be}{\begin{equation}}
\newcommand{\ee}{\end{equation}}
\newcommand{\bea}{\begin{eqnarray}}
\newcommand{\eea}{\end{eqnarray}}

\newcommand{\gapp}{\mathrel{\raise.3ex\hbox{$>$}\mkern-14mu
\lower0.6ex\hbox{$\sim$}}}
\newcommand{\lapp}{\mathrel{\raise.3ex\hbox{$<$}\mkern-14mu
\lower0.6ex\hbox{$\sim$}}}
\def\bbox{{\,\lower0.9pt\vbox{\hrule \hbox{\vrule height 0.2 cm
\hskip 0.2 cm \vrule  height 0.2 cm}\hrule}\,}}

\begin{document}
\title{Separating the superradiant emission from the Hawking radiation from a rotating black hole}
\author{De-Chang Dai$^{1,2}$\footnote{communicating author: De-Chang Dai,\\ email: diedachung@gmail.com\label{fnlabel}}}
\affiliation{ $^1$ Department of Physics, National Dong Hwa University, Hualien, Taiwan, Republic of China}
\affiliation{ $^2$ CERCA, Department of Physics, Case Western Reserve University, Cleveland OH 44106-7079}
\author{Dejan Stojkovic$^{3}$}
\affiliation{ $^3$ HEPCOS, Department of Physics, SUNY at Buffalo, Buffalo, NY 14260-1500, USA}

\begin{abstract}
\widetext
Emission of particles created in the background of a rotating black hole can be greatly amplified taking away rotational energy of a black hole. 
This amplification affects both particles created near the horizon (due to the Hawing effect), and particles created near the potential barrier far from the horizon. Only the latter effect is called the superradiance in the strict sense. We explicitly calculate the superradiant emission for scalar particles and compare it with the total scalar particle emission (Hawking radiation plus superradiance) to clarify some confusion in the literature.  We clearly show that these two emissions are not the same. In particular, superradiance persists even for extremal black holes whose Hawking temperature is zero. 
\end{abstract}


\maketitle

\section{Superradiance}

The notion of superrdiance, or superradiant emission of particles, has been used in a wide range of situations in the literature.
The term "superradiance" was introduced by R.H. Dicke in 1954 \cite{dicke},  describing an effect in which
disordered energy is converted into coherent energy. In classical physics,
superradiance is a classical phenomenon in which an amplitude of an outgoing wave after the reflection is greater
than the amplitude of the ingoing wave \cite{Zeldovich}. This phenomenon can happen in the background of a rotating black
hole \cite{unruh}, which contains an ergosphere, i.e. the region between the infinite redshift surface and the event
horizon.  In such a background, an incident wave can take away some of the rotational energy of the black hole 
and get amplified after reflection, thus effectively yielding a reflection coefficient greater than one (i.e. negative absorption coefficient).

In the context of quantum Hawking radiation from a black hole, it was noticed in numerical studies that spontaneous emission from a black hole can also get amplified taking away
rotational energy of the black hole \cite{Dai:2007ki}. Calculations of the black hole greybody factors indicate that this amplification is very much spin dependent, with emission of
higher spin particles strongly favored. In \cite{Dai:2010xp} an analytic explanation of this phenomenon was given in terms of the spin-spin interaction between the spin of the rotating black hole and spin of the emitted particle.

While the spin dependent amplification of Hawking radiation is also often called superradiance in the spirit of
Dicke's definition in \cite{dicke}, it is different from the original superradiance in \cite{Zeldovich} or \cite{unruh} which crucially rely on the negative absorption coefficient. For example
the superradiance as defined in  \cite{Zeldovich} or \cite{unruh} is not possible for fermions \cite{unruh,Frolov_book}. If the incident wave is
made of fermions, then the reflected wave can not get amplified due to Pauli exclusion principle, since all the
available states are already occupied. In this paper, for clarity, we will call this effect the superradiance and separate it from the Hawking effect. The crucial difference is that the Hawking effect happens in the presence of the horizon, while the superradiance does not need the horizon. Superradiant emission is simply the effect of particle creation in scattering from the potential barrier. 
Since the black hole contains both the horizon and the potential barrier outside the horizon,  the total radiation from the rotating black hole will include both the Hawking effect and superradiance.  

The effects of superradiance in the context of black holes have been extensively explored in the literature (for a comprehensive review see \cite{Brito:2015oca}). However, to the best of our knowledge, it appears that the superradiance has never been explicitly separated from the Hawking radiation in concrete calculations. Therefore, in this paper we will explicitly calculate the particle production due to the potential barrier away from the horizon of a rotating black hole  and compare it with the total radiation. Among the other things, it will become clear that the superradiance exists even when the Hawking temperature drops to zero (i.e. for an extremal balk hole). Thus, treating a black hole as a black body emitter with a finite temperature $T$, and thus intensity of radiation proportional to $T^4$, is a significant oversimplification. This fact can potentially have some implications even for the information loss paradox. Since suprerradiance particles are created at the barrier outside of the horizon, they should not affect the horizon and singularity physics. However, if one takes the whole process of the black hole creation and evaporation into account, potential barrier outside of the horizon still retains some information about the gravitational process that created the black hole. This implies that superradiance is relevant to black hole information\cite{Park:2019lbj}. This is especially important in the context of the black hole scalar hair.    

We start with the metric for a rotating black hole. 
The geometry of a rotating black hole is described by the Kerr metric in Boyer-Lindquist coordinates
\begin{eqnarray}
ds^2 &=& -(1-\frac{2Mr}{\Sigma})dt^2 -\frac{4Mra\sin^2\theta}{\Sigma} dt d\phi +\frac{\Sigma}{\Delta}dr^2 +\Sigma d\theta^2 \nonumber\\
&&+\Big(r^2+a^2 +\frac{2Mra^2\sin^2\theta}{\Sigma}\sin^2\theta \Big) d\phi^2,\\
&&\Delta = r^2 -2Mr +a^2,\\
&&\Sigma=r^2 +a^2\cos^2\theta ,
\end{eqnarray}
 where $a$ is the black hole rotation parameter, while $M$ is the mass of the black hole.
         
\begin{figure}[h]
\includegraphics[width=8cm]{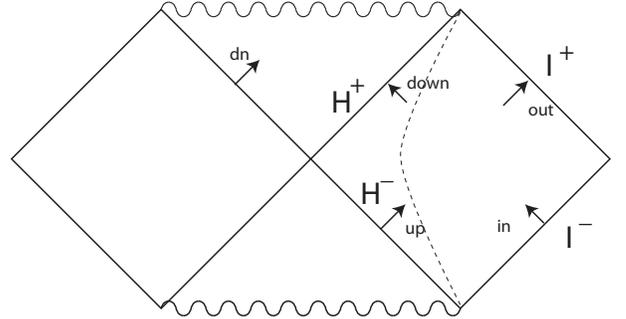}
\caption{The Penrose-Carter diagram: The space outside the horizon is presented in the right square. The upper triangle represents the space inside the future horizon. The black hole radiation due to Hawking effect involves these two regions. On the other hand, only the right square is involved in particle creation by the superradiance mechanism. The thin dashed line represents the potential barrier that induces the superradiance. Four types of bases ($down$, $up$, $in$ and $out$) are involved in the process. Five bases (including $dn$) are involved in the full black hole radiation \cite{Frolov}.        
}
\label{penrose-carter}
\end{figure}

We now place a scalar field in this background. 
We will follow the Frolov's book \cite{Frolov} on black hole physics.  There, a complete classification of different types of bases in the black hole Penrose diagram were given (see Fig.~\ref{penrose-carter}). In this setup one can separate the Hawking effect from superradiance in physically and mathematically clear way. We thus decompose a scalar field $\psi$ in the spherically symmetric coordinates 
\begin{equation}
\psi_{l,m}=e^{-i\omega t}R_{l,m}(r,\omega) \frac{S_{l,m}(\theta,\omega) e^{im\phi}}{\sqrt{2\pi}} ,
\end{equation}
where $l$ and $m$ are angular and magnetic quantum number respectively. The equation of motion can be separated into two main equations, 
\begin{eqnarray}
\label{radius-1}
&&\frac{d}{dr}\Big(\Delta \frac{dR_{l,m}}{dr}\Big) +\Big( \frac{K^2}{\Delta} -\lambda \Big)R_{l,m} =0.\\
&&\frac{1}{\sin\theta}\frac{d}{d\theta} \Big( \sin\theta \frac{dS_{l,m}}{d\theta}\Big) + \Big( a^2\omega^2 \cos^2\theta -\frac{m^2}{\sin^2\theta }\nonumber\\
\label{angular-1}
&&+E\Big)S_{l,m}=0, \\
&&K=(r^2+a^2)\omega -am,\\
&&\lambda=E+a^2\omega^2-2am\omega, 
\end{eqnarray}
 here, $E$ is the eigenvalue of the angular equation, Eq.~(\ref{angular-1}), while $\lambda$ is the corresponding eigenvalue of  the radial equation, Eq.~(\ref{radius-1}). The radial equation can be simplified by introducing a new variable $\chi$,    
\begin{equation}
\chi = (r^2+a^2)^{1/2} R_{l,m} .
\end{equation}
The radial part becomes 
\begin{equation}
\label{radial-part}
\Big(\frac{d^2}{dr_*^2} +\frac{K^2-\lambda \Delta}{(r^2+a^2)^2}-G^2 -\frac{dG}{dr_*}\Big)\chi=0, 
\end{equation}
where, $r^*$ is the tortoise coordinate
\begin{equation}
dr_* =\frac{r^2+a^2}{\Delta} dr .
\end{equation}
and 
\begin{equation}
G=\frac{r\Delta}{(r^2+a^2)^2}
\end{equation}

The asymptotic solution near infinity is 
\begin{equation}
\label{wave-1}
\chi\sim e^{\pm i \omega r_*}, r\rightarrow \infty.
\end{equation}

The asymptotic solution near the horizon is 
\begin{equation}
\label{wave-2}
\chi\sim e^{\pm i (\omega-m\Omega_H) r_*}, r\rightarrow r_+, 
\end{equation}
where, $\Omega_H = a/(2Mr_+)$ and $r_+$ is the radius of the black hole outer horizon.

In the Penrose diagram shown in fig.\ref{penrose-carter}, an incident wave coming from the past infinity, $I^-$, is scattered by the potential. Part of the wave is reflected to infinity, while part of it will penetrate the barrier and fall into the black hole horizon.     
\begin{equation}
 \chi= 
  \begin{cases}
      e^{- i \omega r_*} + \tilde{ R} e^{ i \omega r_*}     & \text{, for }r\rightarrow \infty\\
   \tilde{T} e^{\pm i (\omega-m\Omega_H) r_*}  & \text{, for } r\rightarrow r_+ .
  \end{cases}
\label{wave}
\end{equation}

The radial equation, Eq.~(\ref{radial-part}), is a real second order ordinary differential equation which satisfies the Wronskian relation. From the Wronkian relation, one can prove that 
\begin{equation}
(1-\frac{m\Omega_H}{\omega})|\tilde{T}|^2 = 1-|\tilde{R}|^2 .
\end{equation}
From here we see that if $\omega < m\Omega_H$, then $|\tilde{R}|^2>1$. This means that the reflected wave has a greater amplitude than the initial wave, which in turn implies that particles are created during the scattering. These new particles are not directly created by the black hole horizon, and we focus our attention to this process.

\section{Particle creation by the superradiance mechanism}

As we already explained, the superradiance mechanism operates outside the horizon represented by the line labeled by H$^+$ in fig.\ref{penrose-carter}. So all the relevant events are in the right square in fig. \ref{penrose-carter}. In contrast, Hawking radiation is induced by the black hole horizon.

To study particle creation by superradiance mechanism we have to identify the basis in which we decompose the fields, and the vacuum state of the field.
To describe a vacuum state of a field, we need at least two bases.  For our purpose, we define four possible bases \cite{Frolov}. The first is the $in$-coming mode. It represents a wave which  goes from the past null infinity, $I^-$, to the black hole, 
\begin{equation}
\label{chi-1}
\chi^{in}_{J}\sim\frac{1}{\sqrt{\omega}}\exp(-i\omega r_*) .
\end{equation}
The second one is the $out$-going mode. It represents a wave which  propagates from the black hole to future null infinity, $I^+$, 
\begin{equation}
\label{chi-2}
\chi^{out}_{J}\sim\frac{1}{\sqrt{\omega}}\exp(i\omega r_*)
\end{equation}
The third one is the $down$ mode. It represents a wave which  goes into the future horizon, $H^+$, 
\begin{equation}
\label{chi-3}
\chi^{down}_{J}\sim\frac{1}{\sqrt{|\omega - m\Omega_H|}}\exp(-i(\omega - m\Omega_H) r_*) .
\end{equation}
The forth one is the $up$ mode. It represents a wave which goes away from the past horizon, $H^-$, 
\begin{equation}
\label{chi-4}
\chi^{up}_{H}\sim\frac{1}{\sqrt{|\omega - m\Omega_H|}}\exp(i(\omega - m\Omega_H) r_*) .
\end{equation}

Since a field decomposition requires two distinct bases, we can decompose $\chi$ in two different ways  
\begin{eqnarray}
\hat{\chi} &=& \sum_J \hat{a}_J^{in} \chi^{in}_J+  \hat{a}_J^{ up} \tilde{\chi}^{up}_J+h.c.\\
&=&\sum_J \hat{b}_J^{out} \chi^{out}_J +  \hat{b}_J^{ down} \tilde{\chi}^{down}_J+h.c.
\end{eqnarray}
where, $J=\{\omega,l,m\}$, while $h.c.$ stands for the hermitian conjugate terms. We also have
\begin{eqnarray}
 \tilde{\chi}^{\alpha}_J&= &\chi^{\alpha}_J\text{, if }\omega - m\Omega_H>0\\
 &= &\chi^{\alpha}_J {}^*\text{, if }\omega - m\Omega_H<0 ,
\end{eqnarray}
where $\alpha$ can be $up$ or $down$. The two types of vacuum corresponding to $\hat{a}_{J}^{\alpha}$ and  $\hat{b}_{J}^{\alpha}$ are 
\begin{eqnarray}
\hat{a}_{J}^{\alpha}\ket{in;0} &=& 0\\
\hat{b}_{J}^{\alpha}\ket{out;0} &=& 0 .
\end{eqnarray}

The $in$ mode is expressed in terms of the bases in the past, while the $out$ mode is expressed in terms of the bases in the future. Consider now a field which starts from vacuum in the far past, and after evolving is seen in terms of the $out$ bases
\begin{eqnarray}
\chi^{in}_J&\rightarrow& R_J\chi^{out}_J+T_J \chi^{down}_J \\
\chi^{up}_J&\rightarrow&  t_J\chi^{out}_J+r_J \chi^{down}_J  .
\end{eqnarray}

 If we compare eqs. \eqref{wave-1} and \eqref{wave-2} with eqs. \eqref{chi-1} to \eqref{chi-4}, we see that there are extra normalization factors in the latter four equations. This will lead to extra factors in the transmission coefficients. Therefore,  $T_J$ and $\tilde{T}$ (eq. \ref{wave}) are related by $T_J= \sqrt{|1 - m\Omega_H/\omega|}\tilde{T}$. The creation and annihilation operators are related according to the relationship between the modes. For $\omega - m\Omega_H>0$, we have
\begin{equation}
\hat{b}^{out}_J= R_J\hat{a}^{in}_J+t_J \hat{a}^{up}_J   .
\end{equation}
In this case, there is no particle creation since there is no mixing of the creation and annihilation operators. 
For $\omega - m\Omega_H<0$, we have
\begin{equation}
\hat{b}^{out}_J= R_J\hat{a}^{in}_J+t_J \hat{a}^{up}_J {}^\dagger .
\end{equation}
In this case there is particle creation because of the mixing of the creation and annihilation operators. Thus, $\omega - m\Omega_H<0$ is the necessary condition for the superradiance. From the commutation relations we can get  
\begin{equation}
|t_J|^2= |T_J|^2 .
\end{equation}
From here we can calculate the particle creation number due to the superradiance effect as
\begin{equation}
n_J=\bra{in,0}b_J^{out} {}^\dagger b_J^{out}\ket{in,0}=|t_J|^2 \text{, if }\omega - m\Omega_H<0 .
\end{equation}
To compare the superradiance particle creation with the Hawking effect and demonstrate their difference, we calculate the total  particle creation number (Hawking effect plus superradiance) characterized by the transmission coefficient $|T_J|^2$\cite{Frolov_book}
\begin{equation}
n_J^T=\frac{\text{sign}(1-\frac{m\Omega}{\omega})|T_J|^2}{\exp(\frac{\omega - m\Omega_H}{T})-1}, 
\end{equation}
where, $T$ is the black hole temperature. 

We can immediately see the fundamental difference  between the superradiance and Hawking effect. For example, for an extremal black hole the Hawking temperature goes to zero, $T\rightarrow 0$. If $\omega > m\Omega_H$, which is outside of the superradiant regime, $n_J^T=0$ since both the Hawking effect and superradiance are absent.  However, if $\omega < m\Omega_H$,  we get $n_J=n_J^T$. In that case the Hawking effect is still absent (Hawking temperature is still zero), however the superradiance is present and it is the only contribution to the total radiation from a black hole. 

As an illustration, we will now compare the power spectrum, $\frac{dE}{dt d\omega}$, and power, $\frac{dE}{dt}$, between the superradiance and total radiation.
The power  emitted in particles generated by the superradiance mechanism is 
\begin{equation}
P_s \equiv  \frac{dE_s}{dt} = \frac{1}{2\pi}\int \sum_{l,m} n_J \omega d\omega.
\end{equation}

The total power of particles coming out of the black hole is 
\begin{equation}
P_T \equiv  \frac{dE_T}{dt}= \frac{1}{2\pi}\int \sum_{l,m} n_J^T\omega d\omega.
\end{equation}

\begin{figure}[h]
\includegraphics[width=10cm]{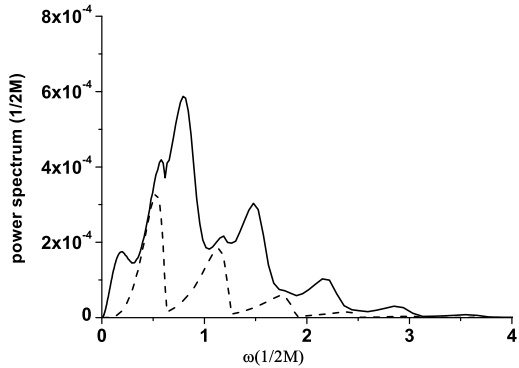}
\caption{  Comparison of the scalar particle power spectra, $\frac{dE}{dt d\omega}$, for $a^*=0.9$. The black curve represents the power spectrum for the total radiation of scalar particles from the black hole (Hawking effect plus superradiance). The dashed line represents  the power spectrum for the scalar particles created by superradiance. The transmission coefficient, $T_J$, is taken from the BlackHawk generator \cite{blackhawk1,blackhawk2}. The spectra are clearly different. We also see that for this value of $a^*$ superradiance is significant but still does not dominate over the Hawking emission.
}
\label{radiation}
\end{figure}

The power spectrum of emitted particles is defined as the emitted energy per unit time per unit frequency, i.e. $\frac{dE}{dt d\omega}$. We use the units where $G=c=k=1$.
In fig. \ref{radiation}, we plot the comparison between the power spectrum of particles created by the superradiance mechanism and the power spectrum of the total radiation from the black hole. For convenience, the rotation parameter $a$ is rescaled to $a^*=a/M$. We fix the value of the black hole rotation parameter to $a^*=0.9$ where the superradiance of scalar particles is significant, but still does not dominate over the Hawking emission.

\begin{figure}[h!]
\includegraphics[width=10cm]{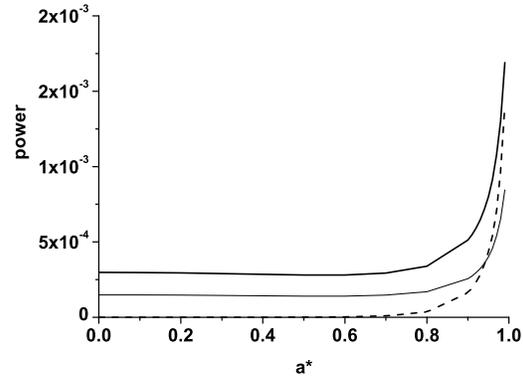}
\caption{  Comparison of the powers for the scalar particles, $P=\frac{dE}{dt}$, as a function of the black hole rotation parameter $a^*$. The black solid curve is the total scalar particle radiation from the black hole (Hawking effect plus superradiance), $P_T$. The thin solid curve is one half of the total radiation, plotted for convenience. The dashed line is the power from the superradiance mechanism, $P_s$. We can see that the superradiance does not play important role for $a^*<0.7$. However, it becomes dominant around $a^*=0.94$. 
}
\label{rate}
\end{figure}

In fig.\ref{rate}, we plot the comparison of the powers as a function of the black hole rotation parameter $a^*$. We can see that the superradiance becomes dominant for highly rotating black holes at $a^* \approx 0.94$.

\begin{figure}[h]
\includegraphics[width=10cm]{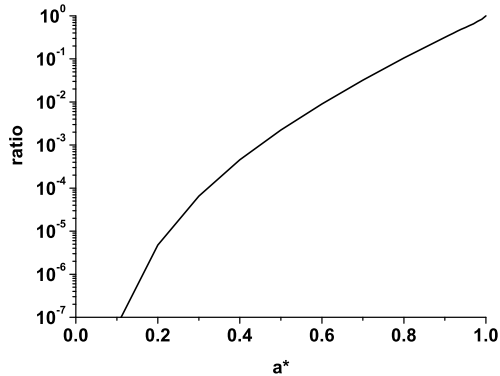}
\caption{ Ratio between the powers emitted by superradiance and total radiation, $P_s/P_T$, for the scalar particles. The curve represents the ratio between the powers of scalar particles created by the superradiance mechanism and the  total radiation (Hawking effect plus superradiance).  We see that at  $a^*=1$ only superradiance contributes to the total radiation. 
}
\label{ratio}
\end{figure}

In fig.\ref{ratio}, we plot the ratio between the powers emitted by superradiance and total radiation.  Once again we see that for the extremal black hole (i.e. $a^*=1$) superradiance equals the total radiation since the Hawking effect ceases to exists at that point. 

\section{Conclusion}
The main point in this paper is to make a pedagogical distinction between the amplification of the Hawking emission in the background of a rotating black hole from the effect of superradiance. 
Both the spin dependent amplification of Hawking radiation and superradiance  crucially depend on taking away rotational energy from a rotating black hole. However, superradiance is created at the potential barrier away from the horizon and should not be mixed with Hawking radiation. We explicitly calculated the superradiant emission and compared it with the total radiation.
 We showed that the superradiance of scalar particles is negligible for $a^*<0.7$, but it becomes dominant around $a^*=0.94$. The numerical values might change if one includes all types of particles since the superradiant amplification is stronger for higher spin particles. 
For the modes that do not satisfy the condition for superradiance, i.e. for  $\omega > m\Omega_H$, we have only the Hawking effect which clearly vanishes for the extremal black hole when the Hawking temperature goes to zero. However, for the superradaint modes, with $\omega < m\Omega_H$, emission persists even for the extremal black  hole. In this paper we considered only a scalar field, but similar calculations can be performed for the vector and graviton fields. As we mentioned in the introduction, fermions do not exhibit superradiance because of the Pauli exclusion principle. 

The fact that even the extremal rotating black holes emit particles might be relevant in various cosmological scenarios that strongly depend on the existence or non-existence of such radiation (e.g.   \cite{Dong:2015yjs,deFreitasPacheco:2020wdg,Chongchitnan:2021ehn,deFreitasPacheco:2023hpb}). In particular, small primordial black holes can Hawking radiate gravitons, thus contributing to the primordial stochastic gravitational wave background  \cite{Dong:2015yjs}.  In our context, even the extremal rotating black holes with the zero Hawking temperature can keep contributing to this background. On the other hand, extremal primordial black holes are often mentioned as good dark matter candidates due to their lack of Hawking radiation  \cite{deFreitasPacheco:2020wdg,Chongchitnan:2021ehn,deFreitasPacheco:2023hpb}. Again, in our context we see that the extremal black holes will keep emitting particles, which makes them unstable and also visible.    

Finally, we comment on the likelihood that small primordial black holes have large angular momentum. Density perturbations usually do not carry significant angular momentum, so the corresponding black holes created by this mechanism will not be spinning fast. However, if two black holes merge, their initial relative angular momentum gets transformed into the final black hole spin, due to the angular momentum conservation. In addition, accretion of surrounding material is very efficient in spinning the black hole up. If the last $20-50 \%$ of the black hole mass came from accretion, such a black hole would be close to extremal. Finally, black holes formed in collisions of the energetic particles in the early universe \cite{Saini:2017tsz} are also expected to carry high spin due to the initial relative angular momentum of the colliding particles.

\begin{acknowledgments}
D.C. Dai is supported by the National Science and Technology Council (under grant no. 111-2112-M-259-016-MY3)
D.S. is partially supported by the US National Science Foundation, under Grant No.  PHY-2014021.  
\end{acknowledgments}

\end{document}